# Monolithically integrated multiple wavelength oscillator on silicon

Jacob S. Levy[1*], Alexander Gondarenko[1*], Mark A. Foster[2], Amy C. Turner-Foster[1], Alexander L. Gaeta[2] & Michal Lipson[1]

[1]School of Electrical and Computer Engineering, [2]School of Applied and Engineering Physics, Cornell University, Ithaca, New York 14853, USA. *These authors contributed equally to this work.

**Silicon photonics enables on-chip ultra-high bandwidth optical communications networks which is critical for the future of microelectronics[1,2]. By encoding information on-chip using multiple wavelength channels through the process of wavelength division multiplexing (WDM), communication bandwidths in excess of 1 Tbit s[-1] are possible[3]. Already several optical components critical to WDM networks have been demonstrated in silicon, however a fully integrated multiple wavelength source capable of driving such a network has not yet been realized. Optical amplification, a necessary component for producing a source, can be achieved in silicon through stimulated Raman scattering[4,5], parametric mixing[6], and the use of silicon nanocrystals[7] or nanopatterned silicon[8]. Losses in most of these previously demonstrated devices have prevented oscillations in those structures. Raman oscillators have been demonstrated[9-11], but the narrow Raman gain window limits operation to a tightly restricted (~ 1 nm) wavelength range and thus is insufficient for WDM. Losses in other previously demonstrated devices have prevented oscillations in those structures. Here we demonstrate the first monolithically integrated CMOS-compatible multiple wavelength source by creating an optical parametric oscillator (OPO) formed by a silicon nitride ring resonator on silicon coupled to an integrated waveguide. The device can generate more than 100 new wavelengths, spaced by a few nm, with operating powers below 50 mW. This source can form the backbone of a fully operational high-bandwidth**

**optical communications network on a microelectronic chip enabling the next generation of multi-core microprocessors.**

Similar to lasing, optical parametric oscillation occurs when the roundtrip parametric gain exceeds the loss in a cavity. Four-wave mixing (FWM) is a third-order nonlinear parametric process, which occurs when two pump photons are converted into a signal photon and an idler photon such that energy is conserved. The efficiency of FWM depends on the pump intensity, the waveguide nonlinearity and interaction length, and the degree of phase mismatch. Phase matching allows the generated light to add up constructively along the length of the waveguide and has been demonstrated in integrated photonic circuits by suitable choice of the waveguide dimensions[6,12,13]. In addition to phase-matching, achieving parametric gain requires relatively high optical intensity and long interaction length. By using a high-quality factor $Q$ optical resonator, the power requirement and device footprint can be reduced due to the photon lifetime and optical field enhancement[14-16] that occurs when all the interacting waves are on cavity resonances. Since the FWM wavelengths are strictly determined by energy conservation, achieving phase-matching in a microresonator allows these cavity resonances to coincide with the generated FWM wavelengths by yielding an equally spaced distribution of resonant modes in energy. Proper design of the ring-waveguide cross-section allows tailoring of the dispersion[12-14] thereby enabling this constant mode spacing. When a pump laser is tuned to a cavity resonance (Fig. 1a), the device is cable of generating simultaneous parametric oscillations at numerous wavelengths. High-$Q$ $CaF_2$ toroidal resonators[17] and silica micro-toroids[18,19] and micro-spheres[20] have shown parametric oscillation based on these principles. However, since these materials possess low nonlinearities, the $Q$ necessary for operation is extremely high. As a result, these devices are sensitive to perturbations and not conducive to on-chip integration since operation requires a purged $N_2$ environment and delicate tapered fiber coupling.

Silicon nitride, the material of choice here for demonstrating an on-chip OPO, has recently been shown to have a nonlinear refractive index, $n_2 = 2.5 \times 10^{-15}$ cm$^2$ W$^{-1}$, about an order of magnitude larger than silica[21]. Silicon nitride is a CMOS compatible material with a linear refractive index of 1.98 at 1550 nm and, due to a larger bandgap, does not suffer from two-photon absorption and the concomitant free-carrier absorption which plagues silicon at communications wavelengths. Deposited silicon nitride films have yielded waveguides with low-losses in both the visible[22,23] and infrared[24,25] regimes. Until recently, the thickness of low-loss silicon nitride waveguides had been restricted to < 250 nm due to tensile stress[24] in the nitride film. Such thin films are poor for nonlinear optics since the waveguide mode is delocalized from the material of interest. Here we grow thicker films using a thermal cycling process described in the methods. A thicker film confines the optical mode to the silicon nitride layer thereby reducing the effective modal area. With control over waveguide thickness, we are able to tailor the waveguide dispersion to achieve phase matching[12,13] of the FWM process. Centering the pump for the FWM process in the anomalous group-velocity dispersion (GVD) regime near the zero-GVD point allows for broad-bandwidth phase-matching and hence signal amplification over a wide wavelength range.

We show that net gain can be achieved in high confinement silicon nitride waveguides. We test a 6.1-cm long waveguide, which has a trapezoidal cross section (Fig. 2c) with a base of 1.45 μm, height of 725 nm, and a wall angle of 23° and is fabricated as described in the methods. According to numerical simulations implemented using a full-vector finite-difference mode solver[13], this geometry should have a zero-GVD point at 1560 nm (Fig. 2b). We measure the propagation loss for the waveguide to be 0.5 dB cm$^{-1}$ with an insertion loss of ~7 dB. To achieve large pump intensities, we modulate a laser centered at 1550 nm at 1 GHz with a 100:1 duty cycle. The modulated light is amplified using a high-power erbium doped fiber amplifier (EDFA). This pump source is combined with a tuneable low power signal using a wavelength-division multiplexer.  The value for the peak pump power $P_{pump}$ in the waveguide is 24 W.  As shown in Fig. 2a, for these conditions we see on/off signal gain over a 150-nm bandwidth and gain as high as 3.65 dB. The total propagation loss through the waveguides is 3 dB, therefore gain above this threshold shows net parametric amplification through the waveguide.

To create an OPO we form a microring resonator using this waveguide design (Fig. 1b).  The measured resonator has a 58-μm radius with a waveguide height of 711 nm, a base-width of 1700 nm, and a sidewall angle of 20° giving anomalous GVD in the C-band and a zero-GVD point at 1610 nm.  The ring is coupled to a single-mode bus waveguide of the same height with a base-width of 900 nm.  The $Q$ of the resonator is 500,000 and the free spectral range (FSR) is 403 GHz (3.2 nm at 1.55 μm wavelength) as measured from the transmission spectrum.  The pump laser is tuned to a resonance near 1557.8 nm.  The power being absorbed by the cavity causes a thermal shift of the resonance.  By slowly tuning the pump frequency deeper into the resonance, we achieve a soft "thermal lock" in which the cavity heating is counteracted by diffusive cooling[26], and the power coupled to the cavity remains constant.  The output spectrum shown in Fig. 3a is measured for a pump power of 310 mW in the waveguide immediately before the ring.  Here we find the generation of 87 new frequencies between 1450 nm and 1750 nm, which corresponds to wavelengths covering the S, C, L and U communications bands.   Design of our source's FSR is accomplished by adjusting the size of the resonator.  For example, using a smaller ring (20-μm radius) with a $Q$ of 100,000, we pump near 1561 nm and generate oscillation in 20 modes of the resonator (Fig. 3b) with modest input powers (150 mW). Using this smaller device, we observe a wider mode spacing of 1.17 THz (9.3 nm at 1.55 μm wavelength).

We measure the linewidth of one of the generated modes by selecting a single mode from the output of our device with a tunable bandpass filter.  Using the technique described in the methods, the measured linewidth of the generated mode is 424 kHz. The measured linewidth of our pump source is 140 kHz. We believe the broadening is caused by the cascaded FWM interactions involved with the generation of the multiple wavelengths and is not strongly influenced by the much broader cavity resonance (~400 MHz). Nevertheless, the demonstrated linewidth is sufficiently narrow for any type of on-chip optical communications. Furthermore, we measure the stability of a single generated frequency by taking a single-shot measurement of the temporal power fluctuations. The measurement shown in Fig. 4a shows less than 5% power variation over 50 μs with a 100-MHz detection bandwidth. By comparing the noise from the generated OPO mode and the background noise on the detector, we calculate the relative intensity noise to have an rms value of 1.6% over this bandwidth range.

The theoretical threshold power for oscillation is proportional to the ring radius and inversely proportional to the quality factor squared[17,27]. In a 40-μm radius ring with a $Q$ of 200,000, we measure the power of the first generated frequency as a function of pump power. We find an oscillation threshold of 50 mW (Fig. 4b). The power within the first oscillating mode saturates when the pump power reaches 100 mW due to cascaded FWM in the ring causing the additional power to be distributed to the newly generated modes. Given that we have recently demonstrated[28] 20-μm radii silicon nitride rings with $Q$'s greater than $10^6$, we estimate a threshold power of less than 5 mW is possible in future devices, which is well within the range of standard laser sources.

We have demonstrated an integrated on-chip multiple wavelength source based on FWM optical parametric oscillation in silicon nitride rings. Using this OPO, numerous equally spaced narrow linewidth sources can be generated simultaneously providing a critical component for realizing a high-bandwidth integrated photonic network that utilizes WDM for communications. Combining this device with previously demonstrated electro-optic components[29] should enable an integrated optical network with bandwidths capable of meeting the interconnect demands for the next generation of microprocessors.

**Figure 1 | On-chip optical parametric oscillator. a,** A single pump laser tuned to the resonance of an integrated silicon nitride microring allows the generation of numerous narrow linewidth sources at precisely defined wavelengths.  This device can dramatically increase the bandwidth of chip-scale communications by encoding information in parallel on these new wavelength channels.  **b,** A scanning electron micrograph of a silicon nitride microring resonator coupled to a bus waveguide.

**Figure 2 | Four-wave mixing in silicon nitride waveguides. a,** Theoretical dispersion of a silicon nitride waveguide showing  zero-GVD at 1560 nm and anomalous dispersion in the C-band.  We pump near the simulated zero-GVD point to maximize our bandwidth.  Inset: scanning electron micrograph of the cross-section of a silicon nitride waveguide depicting the trapezoidal shape of the core and the cladding materials.  **b,** Signal gain as a function of wavelength. On/off gain larger than the 3 dB propagation loss through the waveguide demonstrates a net parametric gain with pump power of 24 W.

**Figure 3 | Demonstration of optical parametric oscillation for integrated multiple wavelength source. a,** The output spectrum of a 58-μm radius silicon nitride ring with a pump wavelength of 1557.8 nm.  The 87 generated wavelengths are equally spaced in frequency with an FSR of about 3.2 nm.  By using each line as a carrier frequency, we can use the source for an on-chip WDM network.  **b,** Using a different ring with a 20-μm radius, we control the frequency spacing of the generated wavelengths.  Here pumping at 1561 nm, we measure 21 wavelengths over a 200-nm span with a spacing of 9.5 nm.

**Figure 4 | Measurement of oscillation stability and threshold. a,** Single shot measurement of power stability in a single oscillating mode. **b,** The output power in the first generated mode compared to the pump power. In this device we find that parametric oscillation occurs with 50 mW of pump power and has a slope efficiency of 2%.

**Methods**

**Device Fabrication.** We use a thermal cycling process to grow thick silicon nitride films that maintain high optical quality. Using a virgin silicon wafer, we grow a 4-μm thick oxide layer for our under cladding. We then deposit between 300-400 nm of silicon nitride using low-pressure chemical vapour deposition (LPCVD). The film is allowed to cool and then annealed for 3 hours at 1200°C in an ambient $N_2$ environment. We deposit a second layer of silicon nitride to bring the total nitride layer to our target thickness. We use MaN-2403 e-beam resist to pattern our waveguides. After exposure and development, we reflow the resist for 5 minutes at 145°C and etch the silicon nitride waveguides using a reactive ion etch of $CHF_3/O_2$. The characteristic trapezoidal cross-section of our waveguides (Fig. 1c) arises from the nature of the resist reflow and etching process. We then clad our waveguides with a thin layer (300 nm) of LPCVD oxide and a thick layer (3 μm) of plasma enhanced chemical vapour deposition (PECVD) oxide.

**Linewidth measurement.** In order to measure the linewidth of a single oscillating mode, we isolate it using a bandpass filter, amplify the filtered signal using an EDFA, and split the signal using a 50/50 coupler. The first arm of the signal is sent through a 20-km spool of fiber so that it is no longer coherent with the other arm. The second arm is modulated at 5 GHz by a pattern generator driving an electro-optic modulator. We recombine the two arms, accounting for the difference in power between them, and send the signal into an electrical spectrum analyzer. By measuring the full-width half-maximum of the noise about the 5-GHz line (ignoring the narrow feature resulting from the unmixed signal), we acquire an accurate measurement of the linewidth.

**Acknowledgements**  The authors would like to acknowledge DARPA for supporting this work under the Optical Arbitrary Waveform Generation Program and the MTO POPS Program.  This work was performed in part at the Cornell NanoScale Facility, a member of the National Nanotechnology Infrastructure Network, which is supported by the National Science Foundation (Grant ECS-0335765).

**Author Information** Correspondence and requests for materials should be addressed to M.L. (ml292@cornell.edu).

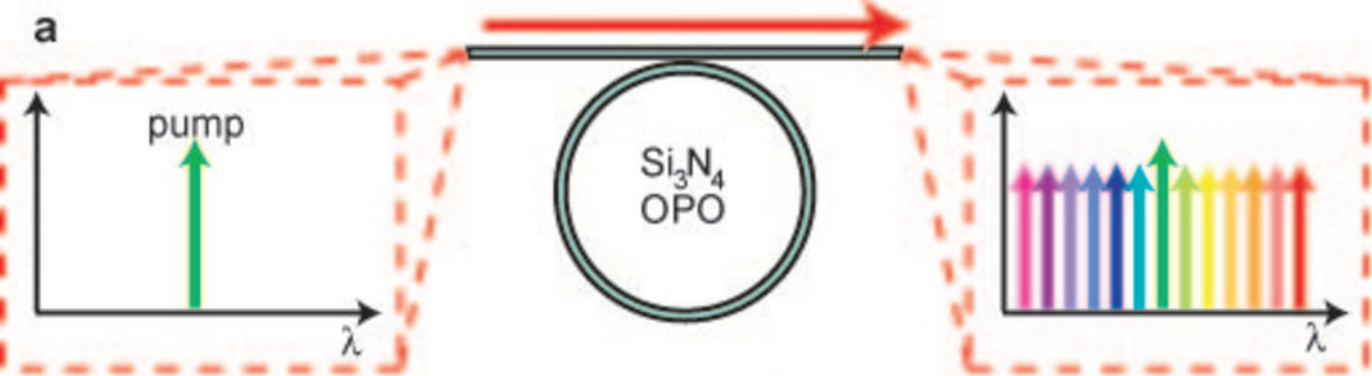

a

pump

λ

Si₃N₄
OPO

λ

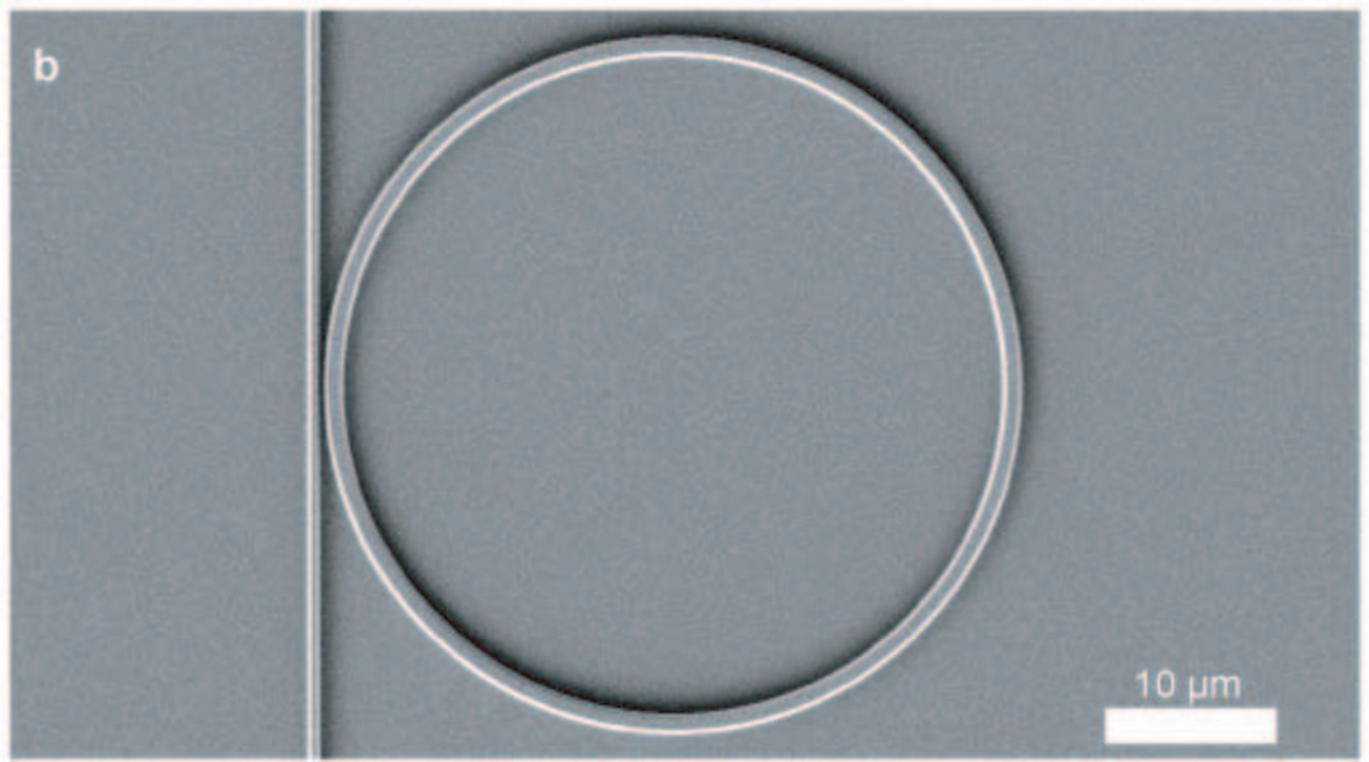

b

10 µm

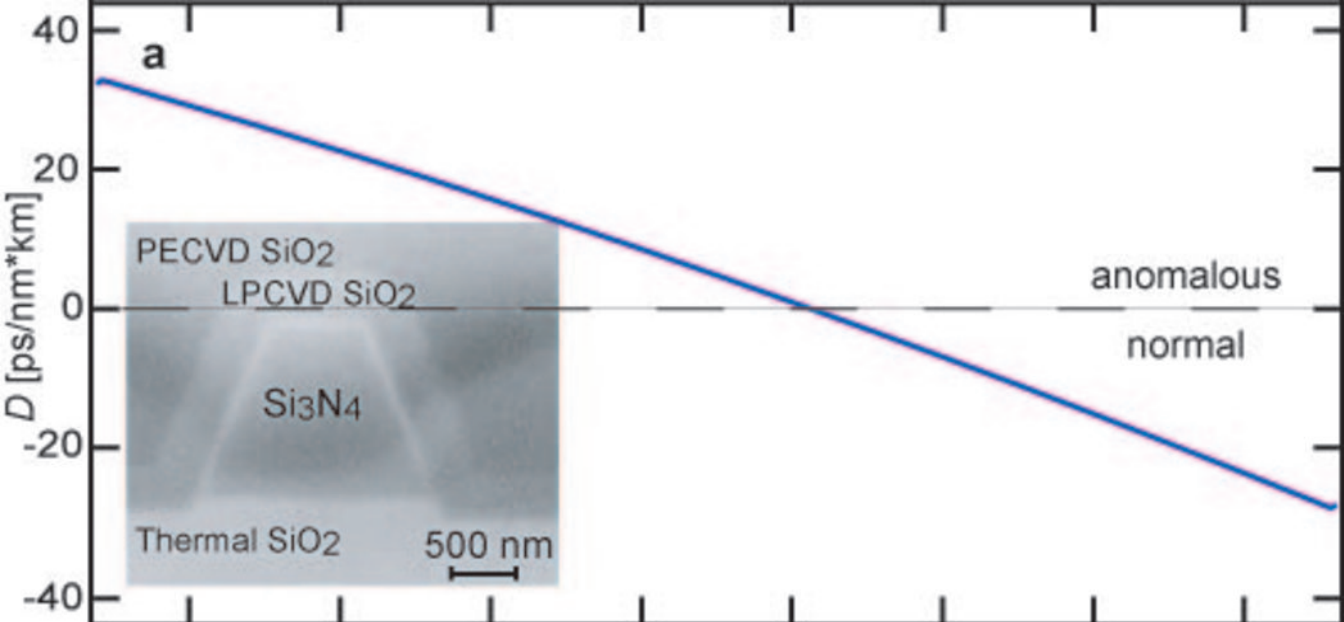

a

PECVD SiO2
LPCVD SiO2
Si3N4
Thermal SiO2
500 nm

anomalous
normal

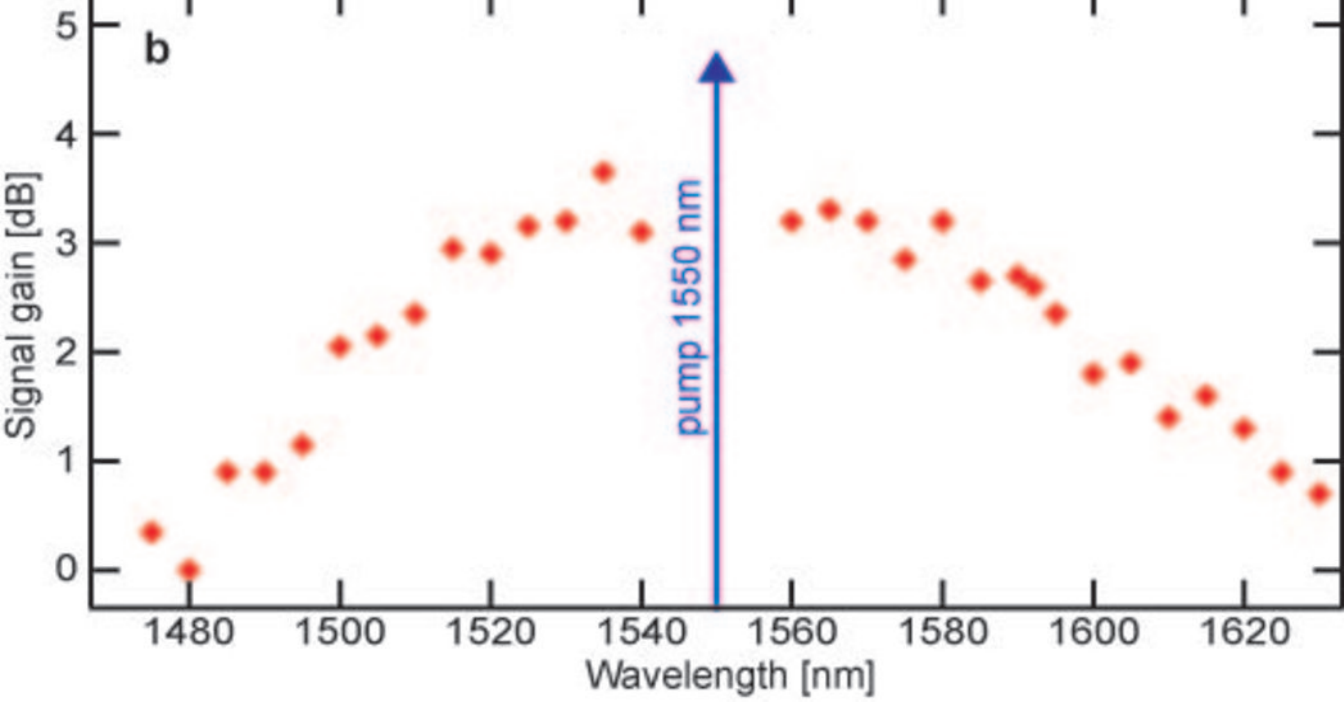

b

pump 1550 nm

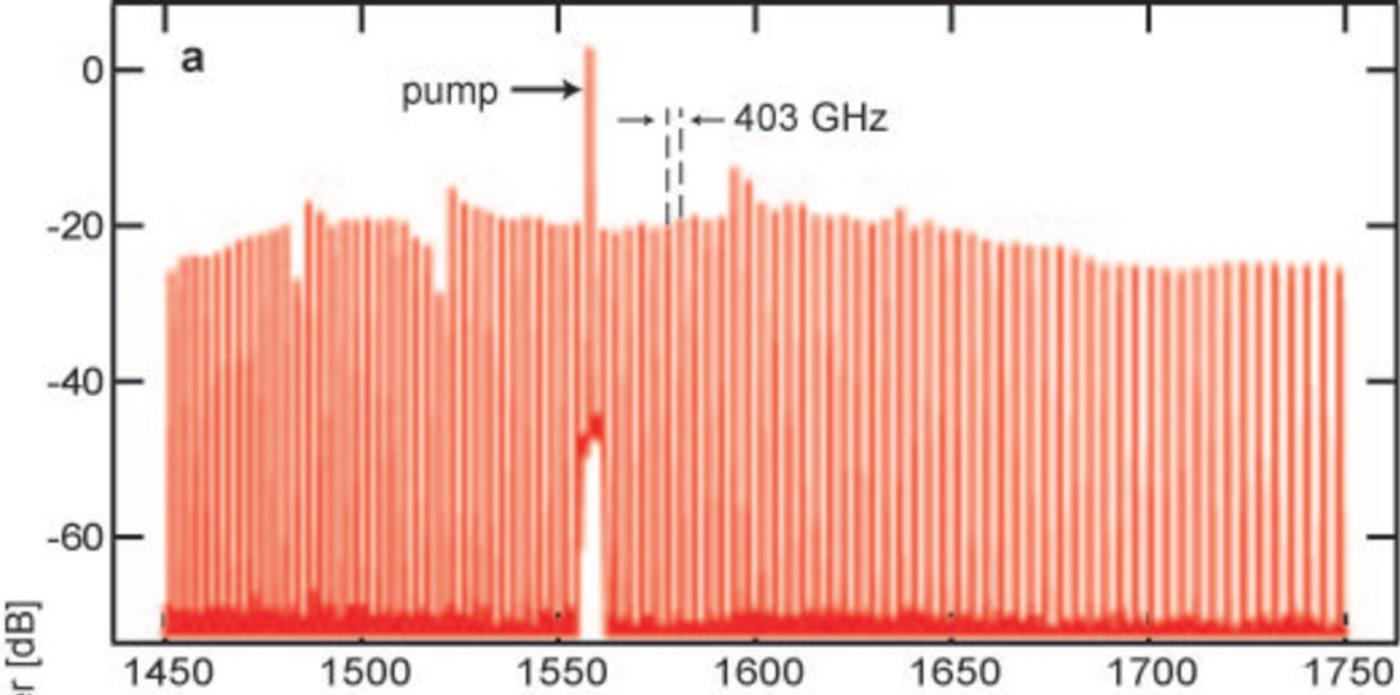

**a**

pump →

→| |← 403 GHz

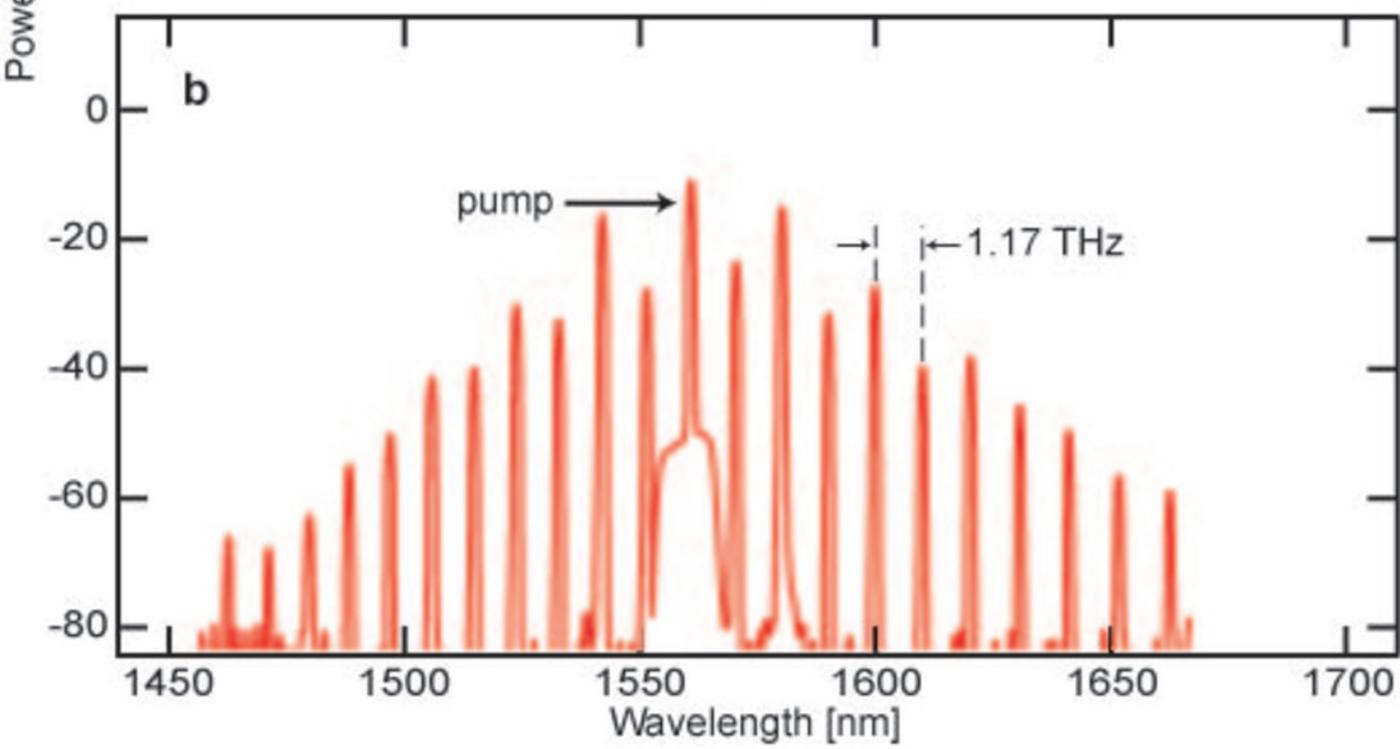

**b**

pump →

→| |← 1.17 THz

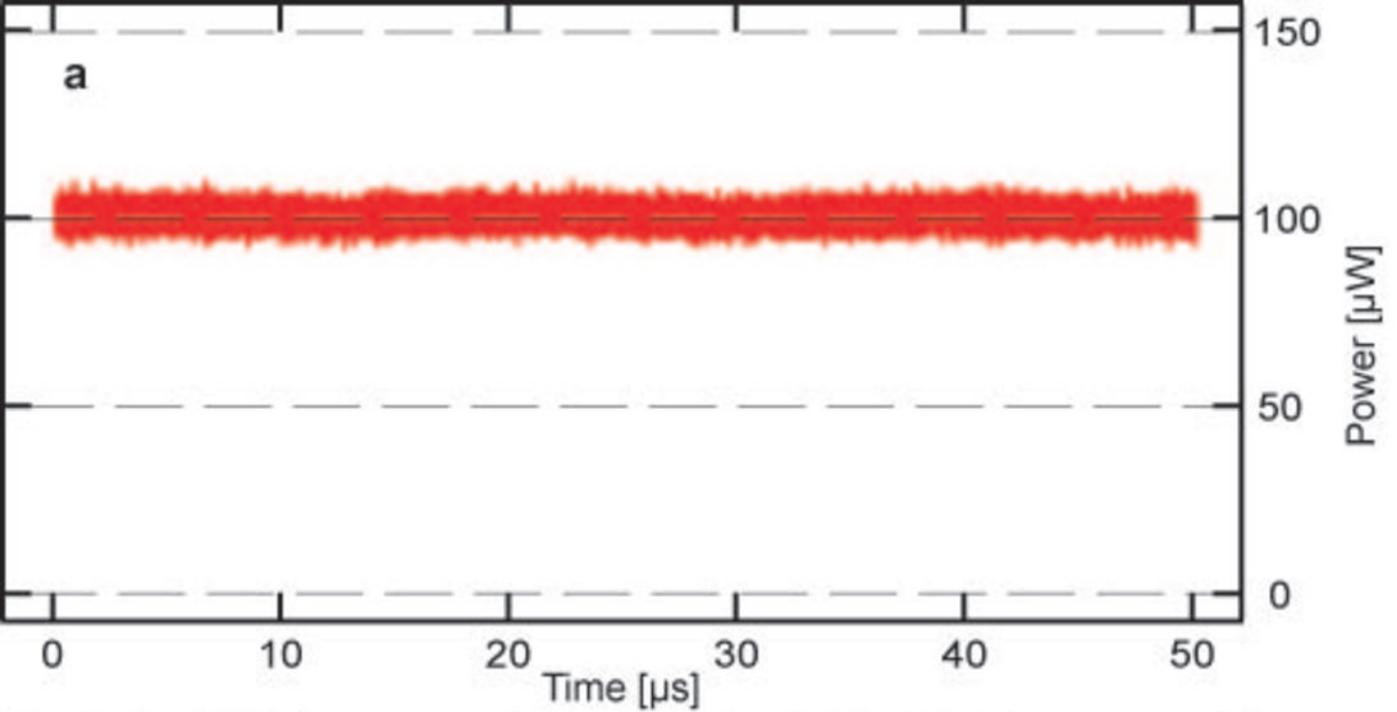

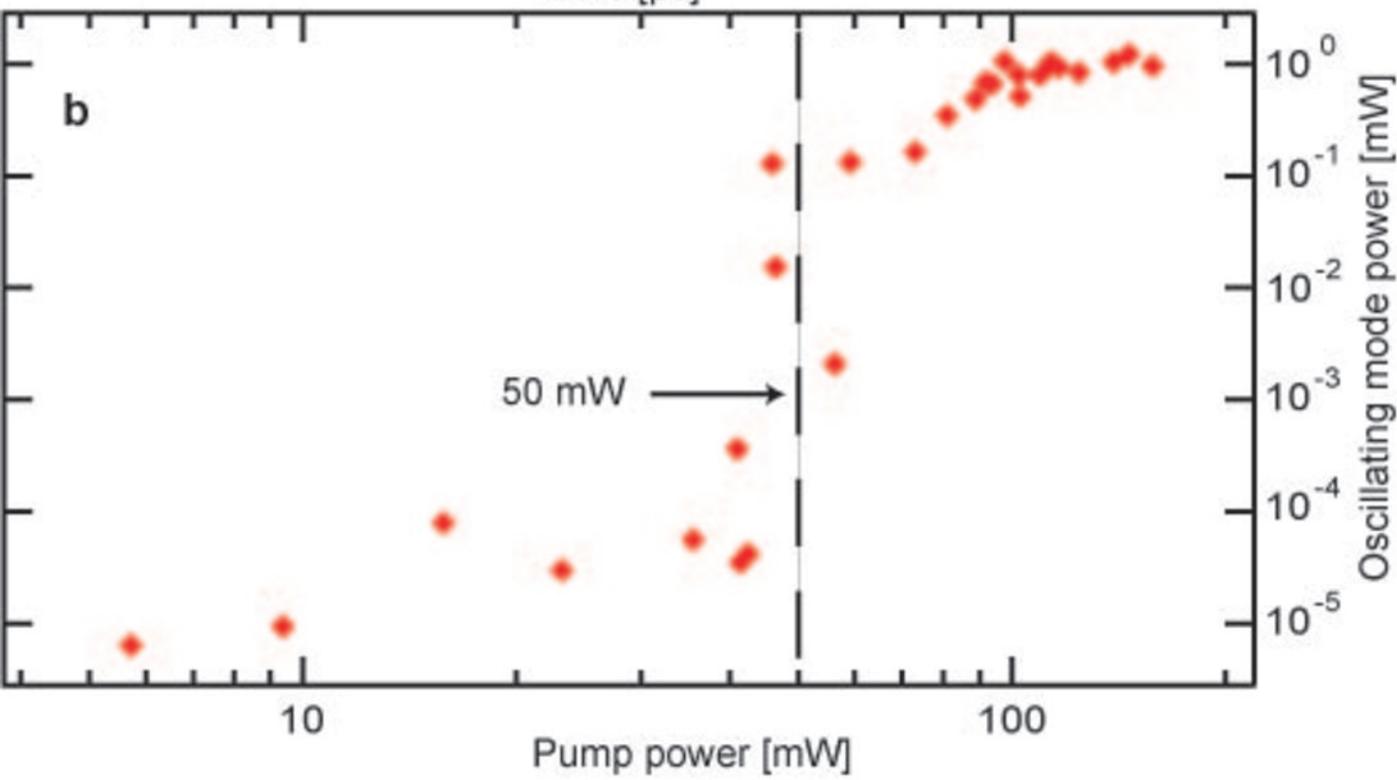